\documentclass[runningheads]{llncs}
\usepackage[T1]{fontenc}
\usepackage[final]{listings}
\usepackage{xcolor} 

\definecolor{LightGray}{rgb}{0.97,0.97,0.97}
\definecolor{Sepia}{rgb}{0.8,0.3,0}
\definecolor{Green}{rgb}{0.0,0.4,0}
\definecolor{Purple}{rgb}{0.5,0,0.9}
\definecolor{MidnightBlue}{rgb}{0.0,0,0.97}

\lstdefinelanguage{ASL}{
  basicstyle=\small\ttfamily,
  backgroundcolor=\color{LightGray},
  columns=fullflexible,
  breaklines=false,
  sensitive=true,
  frame=bt,
  aboveskip=1em,
  belowskip=1em,
  xleftmargin=.5em,
  xrightmargin=.5em,
  framexleftmargin=.5em,
  framextopmargin=.5em,
  framexbottommargin=.5em,
  framexrightmargin=.5em,
  tabsize = 2,
  showstringspaces=false,
  morecomment=[l][\color{gray}]{\#},       
  morecomment=[n][\color{blue}]{<http}{>}, 
  morestring=[b][\color{OliveGreen}]{\"},  
  keywordsprefix=+,
  classoffset=0,
  keywordstyle=\color{Sepia},
  morekeywords={},
  classoffset=1,
  keywordstyle=\color{Purple},
  morekeywords={rdf,rdfs,owl,xsd,purl,phy,req,sys,td},
  classoffset=2,
  keywordstyle=\color{Green},
  morekeywords={eka},
  classoffset=3,
  keywordstyle=\color{MidnightBlue},
  morekeywords={
    SELECT,CONSTRUCT,DESCRIBE,ASK,WHERE,FROM,NAMED,PREFIX,BASE,OPTIONAL,
    FILTER,GRAPH,LIMIT,OFFSET,SERVICE,UNION,EXISTS,NOT,BINDINGS,MINUS,a
  },
  comment=[l]{//},
  morecomment=[s]{/*}{*/},
  commentstyle=\color{purple}\ttfamily,
  stringstyle=\color{red}\ttfamily,
  morestring=[b]',
  morestring=[b]"
}

\usepackage{graphicx}
\usepackage[text=SUBMITTED, fontsize=80pt]{draftwatermark}

\begin{document}
\title{Learnings from Implementation of a BDI Agent-based Battery-less Wireless Sensor}
%
%
\author{Ganesh Ramanathan\inst{1,3}\and
Andres Gomez\inst{2} \and
Simon Mayer\inst{3}}
\authorrunning{Ramanathan et al.}
%
\institute{Siemens AG, Switzerland 
\and TU Braunschweig, Germany \and University of St. Gallen, Switzerland}
\maketitle              
\begin{abstract}
Battery-less embedded devices powered by energy harvesting are increasingly being used in wireless sensing applications.
However, their limited and often uncertain energy availability challenges designing application programs.
%
%
To examine if BDI-based agent programming can address this challenge, we used it for a real-life application involving an environmental sensor that works on energy harvested from ambient light. This yielded the first ever implementation of a BDI agent on a low-power battery-less and energy-harvesting embedded system. Furthermore, it uncovered conceptual integration challenges between embedded systems and BDI-based agent programming that, if overcome, will simplify the deployment of more autonomous systems on low-power devices with non-deterministic energy availability. Specifically, we (1) mapped essential device states to default \textit{internal} beliefs, (2) recognized and addressed the need for beliefs in general to be \textit{short-} or \textit{long-term}, and (3) propose dynamic annotation of intentions with their run-time energy impact.
We show that incorporating these extensions not only simplified the programming but also improved code readability and understanding of its behavior. 

\end{abstract}
\section{Introduction}
Embedded wireless sensors in cyber-physical systems are increasingly adopting energy harvesting as a power source because of the otherwise high cost of wiring or maintenance that would be involved.
However, in energy-harvesting devices (e.g., using photovoltaic sources), the available energy is constrained and can also be nondeterministic. 
Therefore, programmers writing code for controlling the functioning of such devices need to carefully consider, amongst other factors, the energy storage and (expected) harvesting capacity as well as power consumption characteristics of the hardware.

We argue that the design and coding complexity encountered in energy-constrained embedded devices could be reduced through the adoption of a BDI-based agent-oriented programming paradigm~\cite{mas_rao1995bdi}, which has proven benefits in applications that face complex, dynamic, and uncertain environments~\cite{mas_boissier2013multi}.
The availability of a lean BDI framework suitable for embedded devices~\cite{ebdi_william2022increasing,ebdi_dosdirections} and its demonstrated energy efficiency~\cite{ebdi_vachtsevanou2023embedding} makes it possible to apply this approach to energy- and resource-constrained devices in principle.
However, the usage of BDI in industrial practice and by embedded systems programmers today remains elusive due to the missing conceptual bridging of several key aspects.

\begin{figure}[hb]
    \centering
    \includegraphics[width=0.9\linewidth]{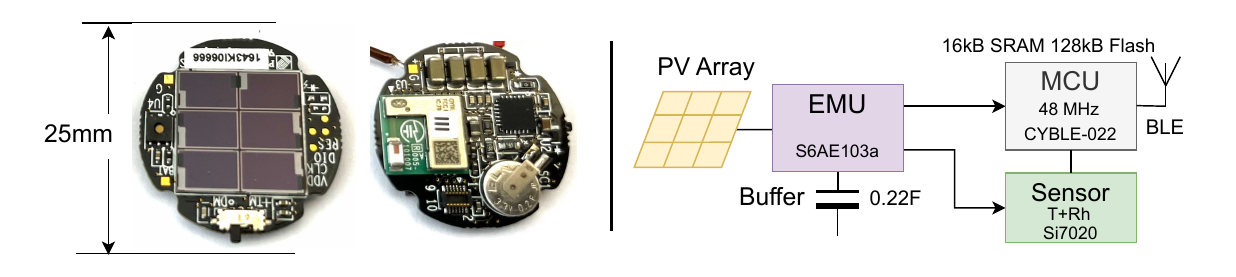}
    \caption{The hardware and a block diagram of the battery-less sensor for which we implemented BDI agent-based energy-conscious application.}
    \label{fig:hardware}
    \vspace{-5mm}
\end{figure}

We implemented a BDI agent-based control program for a real battery-less embedded hardware platform that is powered by energy-harvesting photovoltaic cells (see Figure~\ref{fig:hardware}).
Based on the experience gathered from our implementation, we share three vital design-related insights that lie in the confluence of energy-constrained embedded systems and BDI-style programming: First, we identified key states of the device which play an important role in managing its energy-efficient operation and recommend them to be modeled as \textit{default internal beliefs}. These internal beliefs then aid in the contextualization of the plans in a library and direct the developer's attention towards handling changes in them (e.g., through explicit event handlers). 
Second, we discuss ways through which the energy impact of intents and their contained actions can be included as metadata that enables evaluating the feasibility of achieving goals at run time. We show that the agent can also update this information as it \textit{learns from experience} about these energy impacts.
Third, because energy-constrained devices are often designed to enter into \textit{deep-sleep} modes where they power down their RAM, we propose that beliefs, which serve to model the mental state of the agent, be flagged as being \textit{short-} or \textit{long-termed} so that the state of the program can be persisted and restored correctly when the device switches back to active mode. 

Together, our propositions lead to a significant and highly relevant alignment of the considerations of embedded systems developers with agent-oriented programming, and specifically with the BDI framework.
After a brief overview of related work, we detail our approach in Section~\ref{sec:approach}. In Section~\ref{sec:eval}, we discuss our evaluation of these proposals in the scope of a re-implementation of a control program for a constrained embedded device, which shows a clear qualitative improvement in the design (e.g., concerning the clearer contextualization and legibility of intentions) without disadvantaging energy efficiency. 

\section{Related Work}
Achieving optimal control of low-power and energy-constrained embedded devices is a widely researched topic~\cite{bat_gomez2018design}.
Frameworks and operating systems like Contiki~\cite{low_Contiki-NG} provide infrastructural support for programming resource-constrained devices through libraries of functions for managing memory, event handling, and communication in an energy-efficient and resource-conscious manner.
However, the design of the application logic itself can often become complex when the available energy is not deterministic~\cite{bat_gomez2018design}.
For such circumstances, BDI-style agent programming offers benefits because the plans in its library could be formulated and contextualized according to energy-related aspects. 
In~\cite{ebdi_vachtsevanou2023embedding}, the Embedded BDI framework~\cite{ebdi_william2022increasing} was used to program low-power sensors.
Extending this work, to the best our knowledge, our research is the first to demonstrate a successful implementation of BDI agents on an embedded system which is low-power, energy-harvesting, and battery-less.
We use our implementation to highlight some vital design integration points between such embedded systems and BDI-based agent programming.

\section{Approach}
\label{sec:approach}
We based our battery-less sensor to observe room temperature and humidity on a hardware platform\footnote{\url{https://www.infineon.com/cms/en/product/evaluation-boards/cyalkit-e02/}} that is powered by an onboard energy-harvesting system (see Figure \ref{fig:hardware} for details).
We implemented the control logic for this sensor using the Embedded BDI framework.
We find that the use of the BDI model in energy-constrained devices is beneficial because the selection of a control strategy (which ultimately realizes the agent's intentions) can be \textit{explicitly and visibly} contextualized in the current state of the device (e.g., the energy available in the buffer), goal priorities, and concurrent desires (e.g., to conserve energy) while simultaneously being reactive to changes (e.g., in available energy or network state). 

After the first iteration of implementation, we examined the application program and identified three key aspects that played a significant role in designing the control logic.
We now present these findings by abstracting away the implementation-specific details and instead focusing on the high-level design-related learnings that we believe are more widely valid (and interesting) for agent-oriented programming for resource-constrained devices overall.

\textbf{\textit{Energy impact} of intents:} A control program that operates unaware of energy constraints risks emptying the buffer to the extent that it causes the CPU to lose power unexpectedly.
Therefore, the expected energy consumption of actions such as using a peripheral device or the radio is estimated upfront using the hardware design specification, and the program functions are designed accordingly.
However, the actual consumption may vary at run time depending on the exact operational conditions.
For such cases, a control program can potentially access the EMU to retrieve and \textit{learn} the energy impact. 
For this purpose, we modeled the energy requirement of each intent as an associated belief. We followed the convention that these beliefs carry the names of their associated plans, with a prefixed \texttt{e\_}.
These beliefs can either be set at compile time (based on estimates) or dynamically updated at run time before and after the intent has been executed.
As shown in Listing~\ref{lst:example-asl}, this helps the agent choose a plan which can be accomplished given the current state of the energy buffer.

\begin{lstlisting}[language=ASL,numbers=left, caption={Example code showing how an intents are contextualized by energy-related beliefs (which in turn, are enriched by annotations pertaining to the platform)}, label={lst:example-asl}]
e_meas_temperature(30)[persist("fram")]. //energy estimate (uJ)
e_available(0). //Energy (uJ) available for the application 
e_tendency(0). //Rate of change in input energy (uJ/hr)
transmit_power(8)[impact(101)]. //8dBm needs 101mJ
transmit_power(4)[impact(30)]. //4dBm needs 30mJ

+!broadcast(A):transmit_power(P)[impact(E)] & A > E
<- start_ble_adv(P). //Tx power is chosen based on available energy

+!meas_temperature: e_available(A) & e_meas_temperature(R) & A > R
<-  energy_checkpoint();
    read_trh_sensor();
    !transmit_data;
    update_estimate("e_meas_temperature");
    deep_sleep().
    
+!transmit_data: e_available(A) & e_tendency(I) & I > 50 
<-  !broadcast(A - 50). //50mJ is held as reserve

+!transmit_data // Aggregate and send later
<-  store_for_later_tx().
\end{lstlisting}

\textbf{Incorporating energy-related states as \textit{internal beliefs}:} Plans are often contextualized by the state of the embedded device. Specifically, information about the state of the Embedded Microcontroller Unit (EMU) appears to be necessary in most cases.
We examined our implementation to identify states that played a role in designing the energy-efficient control of the sensor and grouped them into broad categories (see Table~\ref{tab:states}).
As a general design suggestion, we propose that agent programmers (or even embedded systems designers) examine the availability and relevance of device states in these categories to determine if they can play a role in contextualizing the plans and, if so, model them as \textit{beliefs}. 
Further, since some device states such as network role have a relevant impact on energy consumption, the feasibility to annotating beliefs with their estimated power consumption is helpful to contextualize plans in a more granular manner. Lines 4-8 in Listing~\ref{lst:example-asl} illustrate this possibility.

\begin{table}
\caption{Relevant device states for determining energy-conscious agent behavior.}
\label{tab:states}
\begin{tabular}{|l|l|}
\hline
\textbf{State}        & \textbf{Use in control logic} \\ \hline
Device mode           & Entering active mode signals agent to add measurement goal\\ \hline
Network role          & Remain in BLE peripheral role when energy is low\\ \hline
Network state         & Update BLE advertisement payload only when the radio is initialized \\ \hline
Buffer size           & To determine the advertisement interval \\ \hline
Buffer state          & Adapt BLE transmit power to conserve energy  \\ \hline
Buffer input rate     & Reduce advertisement interval when ambient light decreases \\ \hline
\end{tabular}
\end{table}

\textbf{Distinguishing the \textit{lifetime of beliefs}:} Low-power embedded platforms often support something known as the \textit{deep-sleep} mode where the RAM and most peripherals are powered down to save energy.
Before entering this mode, applications can store required data in non-volatile memory. However, this has to be used with constraint since accessing it impacts energy consumption. Furthermore, multiple different options for non-volatile storage (e.g., Flash as well as FRAM) may be available on the same device.
To enable distinguishing beliefs that should be persisted, we propose to annotate beliefs with a custom field \texttt{persist(none|fram|flash)}. 
We extended the cross-compiler to create two function skeletons \texttt{persist} and \texttt{restore}, which are called with a list of variables (and their storage location) when the device mode changes.

\section{Evaluation}
\label{sec:eval}
In the second iteration of our implementation, we systematically considered the design aspects described in the preceding section.
The broad categories of the device state we identified as relevant to contextualizing the plans served as a helpful guideline during programming.
A more practical adoption of this approach could be facilitated by a tool that automatically extracts such beliefs from variables flagged in the hardware abstraction layer of the embedded system.

The platform-agnostic assignment of beliefs as long- or short-term can be achieved through annotations; beyond the advantages this brings for embedded BDI, such annotations could furthermore be used to qualify the use of a belief when contextualizing a plan: for instance, a plan may only rely on beliefs that are long-term.
By quantifying the energy impact of each intent, we can ensure that the agent never chooses an intent that is expected to cause the device to power down unexpectedly. This approach also simplifies the introduction of alternative plans, potentially leading to a more efficient utilization of energy.

Our initial comparisons of the energy consumption of the BDI-based implementation with a traditional C program did not reveal a large difference; however, further tests are required to quantify this.

In conclusion, our findings demonstrate that infusing agent programs with energy-relevant knowledge is highly beneficial for creating BDI-based software for energy-conscious settings. This approach facilitates the construction of control logic that achieves higher utilization of the available energy and addresses challenges from the conflation of energy-related aspects and functional aspects of the system, a significant advantage in energy-conscious settings.

\bibliographystyle{splncs04}
\bibliography{references}

\begin{thebibliography}{1}
\providecommand{\url}[1]{\texttt{#1}}
\providecommand{\urlprefix}{URL }
\providecommand{\doi}[1]{https://doi.org/#1}

\bibitem{mas_boissier2013multi}
Boissier, O., et~al: Multi-agent oriented programming with jacamo. Science of Computer Programming  \textbf{78}(6),  747--761 (2013)

\bibitem{bat_gomez2018design}
Gomez, A.: Design and Specification of Batteryless Sensing Systems. Ph.D. thesis, ETH Zurich (2018)

\bibitem{low_Contiki-NG}
Oikonomou, G., et~al: The {Contiki-NG} open source operating system for next generation {IoT} devices. SoftwareX  \textbf{18},  101089 (2022). \doi{https://doi.org/10.1016/j.softx.2022.101089}

\bibitem{mas_rao1995bdi}
Rao, A.S., Georgeff, M.P., et~al.: Bdi agents: from theory to practice. In: Icmas. vol.~95, pp. 312--319 (1995)

\bibitem{ebdi_dosdirections}
dos Santos, M.M., H{\"u}bner, J.F., de~Brito, M.: Directions for implementing bdi agents in embedded systems with limited hardware resources

\bibitem{ebdi_vachtsevanou2023embedding}
Vachtsevanou, D., et~al: Embedding autonomous agents into low-power wireless sensor networks. In: International Conference on Practical Applications of Agents and Multi-Agent Systems. pp. 375--387. Springer (2023)

\bibitem{ebdi_william2022increasing}
William, J., et~al: Increasing the intelligence of low-power sensors with autonomous agents. In: Proceedings of the 20th ACM Conference on Embedded Networked Sensor Systems. pp. 994--999 (2022)

\end{thebibliography}

\end{document}